\documentclass[aps,superscriptaddress,twocolumn,noshowpacs]{revtex4}
\usepackage[latin1]{inputenc}
\usepackage{graphicx}
\usepackage{amsmath,amssymb}
\usepackage{dcolumn}
\usepackage{bm}
\usepackage{longtable}

\newcommand{\kpe}{\mathbf{k}\!\cdot\!\mathbf{p}}

\newcommand{\tn}[1]{\textnormal{#1}}

\allowdisplaybreaks
\begin{document}

%\preprint{HEP/123-qed}

\title{Polarized emission lines from $A$- and $B$-type excitonic complexes in single InGaN/GaN quantum dots}

\author{M.~Winkelnkemper}
	\email{momme@sol.physik.tu-berlin.de}
\affiliation{
Institut f\"ur Festk\"orperphysik, Technische Universit\"at Berlin, D-10623 Berlin, Germany}
\affiliation{Fritz-Haber-Institut der Max-Planck-Gesellschaft, D-14195 Berlin, Germany}

\author{R.~Seguin} 
\affiliation{
Institut f\"ur Festk\"orperphysik, Technische Universit\"at Berlin, D-10623 Berlin, Germany}
\author{S.~Rodt} 
\affiliation{
Institut f\"ur Festk\"orperphysik, Technische Universit\"at Berlin, D-10623 Berlin, Germany}
\author{A.~Schliwa}
\affiliation{
Institut f\"ur Festk\"orperphysik, Technische Universit\"at Berlin, D-10623 Berlin, Germany}
\author{L.~Rei{\ss}mann} 
\affiliation{
Institut f\"ur Festk\"orperphysik, Technische Universit\"at Berlin, D-10623 Berlin, Germany}
\author{A.~Strittmatter} 
\affiliation{
Institut f\"ur Festk\"orperphysik, Technische Universit\"at Berlin, D-10623 Berlin, Germany}
\author{A.~Hoffmann} 
\affiliation{
Institut f\"ur Festk\"orperphysik, Technische Universit\"at Berlin, D-10623 Berlin, Germany}
\author{D.~Bimberg} 
\affiliation{
Institut f\"ur Festk\"orperphysik, Technische Universit\"at Berlin, D-10623 Berlin, Germany}

\date{\today}

\begin{abstract}
Cathodoluminescence measurements on single InGaN/GaN quantum dots (QDs) are reported. Complex spectra with up to five emission lines per QD are observed. The lines are polarized along the orthogonal crystal directions [$11\overline{2}0$] and [$\overline{1}100$]. Realistic eight-band $\kpe$ electronic structure calculations show that the polarization of the lines can be explained by excitonic recombinations involving hole states which are either formed by the $A$ or the $B$ valence band. 
\end{abstract}


\maketitle

Nitride-based semiconductor nanostructures have been widely studied in the last decade and revolutionary opto-electronic devices have emerged from this effort. Ultra-bright light emitting diodes and laser diodes in the blue spectral region have been realized based on InGaN structures \cite{nakamurabook}. Single-photon emission from GaN/AlN quantum dots (QDs) has been demonstrated recently \cite{santori2005,kako2006}. 
However, the electronic structure of nitride-based QDs is yet poorly understood.  Emission lines from confined excitons and biexcitons in InGaN QDs have been identified; their binding energies, however, scatter across a wide range \cite{seguin2004,rice2005,schoemig2004,sebald2006}. We demonstrate in this Paper that up to five spectrally narrow lines from one and the same QD can be found.  The lines show a pronounced linear polarization, which could be exploited for the implementation of quantum key distribution protocols \cite{BB84} using single photon emitters.  We will show that the polarization of the lines is owed to the valence band (VB) structure of wurtzite (wz) group-III nitrides and the specific strain distribution within QDs grown on (0001).  Realistic eight-band $\kpe$ modeling of the electronic structure reveals, that an electron in the ground state ($e_0$) has a substantial probability to recombine with a hole in either the ground state, which is formed by the $A$ VB ($h_0 \equiv h_A$), or the first excited state, which is formed by the $B$ VB ($h_1 \equiv h_B$).  The different characters of the $A$ and $B$ VBs lead to orthogonal polarization directions of both transitions if a slight structural anisotropy of the QD is present. 
The observed polarizations can thus be explained by recombinations of confined excitonic complexes involving holes in the $A$ or $B$ hole states. Examples for such complexes are the $A$ exciton ($X_A$, with the hole occupying $h_A$), $B$ excitons ($X_B$, with the hole occupying $h_B$), or higher excitonic complexes involving $h_A$ or $h_B$.  Biexcitons with a mixed configuration ($XX_{AB}$), i.e., with one hole occupying $h_A$ and one occupying $h_B$, have already been observed in bulk GaN \cite{adachi2003}.

The sample investigated in this work was grown on Si(111) substrate by low-pressure metal-organic chemical vapor deposition using a horizontal AIX200 RF reactor. An AlAs layer was grown and subsequently converted to AlN as a nucleation surface \cite{strittmatter1999}. In the following step an Al$_{0.05}$Ga$_{0.95}$N/GaN buffer layer was grown at T=1150\,$^{\circ}$C up to a total thickness of 1\,$\mu$m.  The InGaN layer was grown at 800\,$^{\circ}$C with a nominal thickness of 2\,nm using trimethylgallium, trimethylindium, and ammonia as precursors. The QDs form within the InGaN-layer due to spinodal decomposition.  The growth was finished with a 20\,nm GaN cap layer grown during the heat-up phase to 1100\,$^{\circ}$C.

The sample was investigated using a JEOL JSM 840 scanning electron microscope equipped with a cathodoluminescence setup \cite{bimberg1985}. It was mounted onto a He-flow cryostat providing temperatures as low as 6\,K. The luminescence light was dispersed by a 0.3\,m monochromator with a 2400 lines/mm grating and detected with a nitrogen-cooled Si-charge-coupled-device camera, giving a spectral resolution of 310\,$\mu$eV at 3\,eV. In order to increase the spatial resolution we applied Pt shadow masks onto the sample surface with aperture diameters of 200\,nm \cite{tuerck2000}.  The light emission perpendicular to the sample surface ($\mathbf{k}\parallel c$-axis) is detected.  For the determination each line's polarization direction a polarization filter in the detection path was rotated and the corresponding spectra were recorded. The angle of maximum light emission was determined by fitting a $\mathrm{cos}^2$ formula to the intensities of the single lines as a function of the polarizer's angle. The angles were mapped to crystal directions based on the substrate orientation as given by the supplier of the Si substrates.
\begin{figure}
  \includegraphics[clip,width=.7\columnwidth]{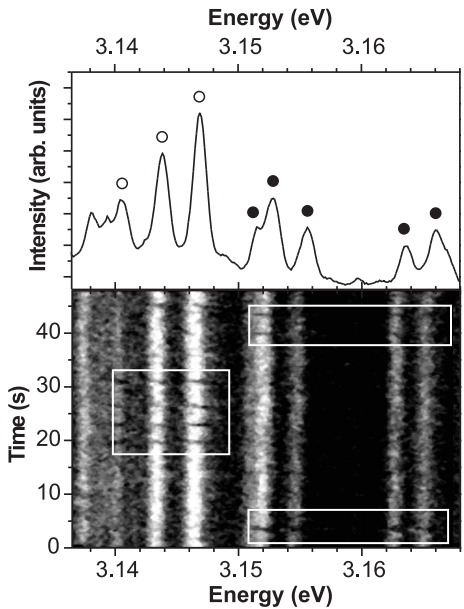}
  \caption{Temporal evolution of a typical InGaN QD spectrum.  The intensity is coded in gray scale. The series consists of 160 spectra, each being integrated for 300\,ms.  Two groups of lines that show the same jitter are marked by the full and empty dots.  The white rectangles highlight characteristic kinks in the jitter traces.}
\label{figure1}\end{figure}

The integral CL spectrum at 6\,$^{\circ}$K shows an intense peak centered at 2.98\,eV with a full width at half maximum (FWHM) of 80\,meV originating from the InGaN layer. When measured through one of the apertures the \mbox{InGaN} peak decomposes into sharp lines. The lines show a FWHM of less than 1\,meV, the narrowest of 0.48\,meV. These lines can be found over a wide energy range from $2.8$ to $3.2$\,eV. Since the line density is very high we investigated the high energy side of the ensemble peak where the single lines are well resolved.  The individual lines are not stable, but vary over a timescale of some 100\,ms in their energetic positions and intensities.

This jitter is caused by variations of the local electric field at each QD's position. Hence, all lines originating from the same QD show the same jitter pattern \cite{tuerck2000,seguin2004,rodt2005,bardoux2006}.  An example is shown in Fig.~1.  Line groups originating from one and the same QD have been successfully identified in other material systems (InAs/GaAs \cite{rodt2005}, CdSe/ZnSe \cite{tuerck2000}, GaN/AlN \cite{bardoux2006}) using these characteristic jitter patterns. 
\begin{figure} 
  \includegraphics[clip,width=.9\columnwidth]{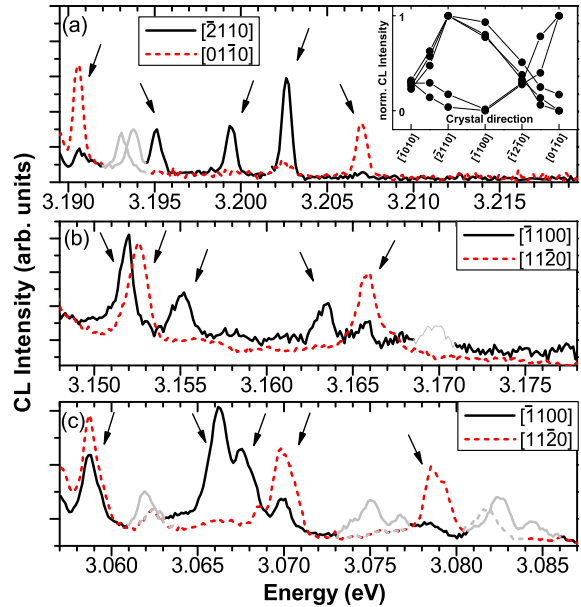}
  \caption{Polarized spectra of three different QDs are shown. Arrows indicate lines displaying the same jitter and belonging to the same QD. Gray lines originate from other QDs.  The inset shows the evolution of the peak intensities as a function of the polarization angle.  Closer analyses yield deviations from the [$11\overline{2}0$] and [$\overline{1}100$] crystal directions of $13\,^{\circ}\pm16\,^{\circ}$ for QD (a), $-4\,^{\circ}\pm4\,^{\circ}$ for QD (b), and $-13\,^{\circ}\pm2\,^{\circ}$ for QD (c).}
\label{figure2}\end{figure}
Groups of up to five lines displaying the same jitter could be found, indicating the existence of a number of different excitonic complexes in the same QD. (Fig.~2)  The typical energetic spread of the five lines was found to be $< 20$\,meV.  All lines show a pronounced linear polarization in orthogonal directions.  The polarization directions scatter around the [$11\overline{2}0$] and [$\overline{1}100$] direction. Both directions were found in each investigated line group. (Fig.~3)  
\begin{figure}
  \includegraphics[clip,width=.9\columnwidth]{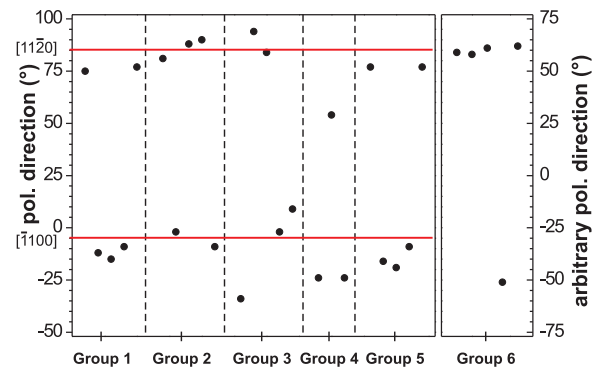}
  \caption{Polarization directions of six different line groups.  They scatter around [$11\overline{2}0$] and [$\overline{1}100$].  Both directions can be found in each line group. For line group No.~6, the alignment of the polarization directions to the crystal axes is unknown. Groups 1-3 correspond to the spectra (a)-(c) in Fig.~2.\label{figure3}}
\end{figure}
Such behavior has not been observed before in nitrides and deserves closer investigation. In III-V arsenides, phosphides and II-VI materials polarized emission lines from QDs have been assigned to the fine structure of the exciton \cite{gammon1996, seguin2005,bayer2002,sugisaki1999,kulakovskii1999} or charged biexciton emission \cite{akimov2005, seguin2006}. The source for both effects is the anisotropic exchange interaction between confined electrons and holes, which leads to a systematic pattern of the polarized lines (polarized doublets, similar order or similar energetic distance of co-polarized lines for different QDs). No such scheme can be deduced from the recorded spectra here (Fig.~2).

We hence suggest a different mechanism to be responsible for the polarization of the lines.

\begin{figure} 
  \includegraphics[clip,width=1.0\columnwidth]{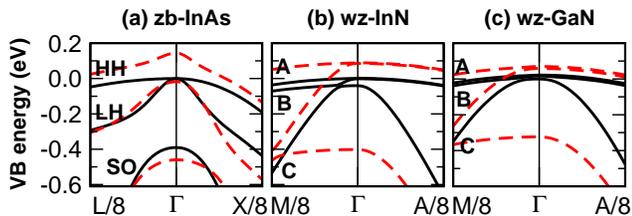}
  \caption{
        Calculated bulk $\kpe$ VB structure of (a) zb-InAs, (b) wz-InN, and (c) wz-GaN; unstrained (black solid lines) and under compressive biaxial strain (-3\,\%) in the basal plane with elastic relaxation along (001) direction (red dashed lines).  The splitting between the $A$ and $B$ VB at $\Gamma$ is $\approx 3$\,meV in strained and unstrained InN and $11$\,meV ($5$\,meV) in strained (unstrained) GaN.
       \label{figure4}
       } 
\end{figure}

In QDs with larger lattice constants than the surrounding matrix the dominant strain components are compressive hydrostatic strain and compressive biaxial strain in the growth plane. The biaxial strain modifies the VB structure:  In zinc blende (zb) InAs QDs it leads to a splitting of the two top VBs---the heavy hole ($HH$) and the light hole ($LH$) band---resulting in a clear separation of all three VBs ($HH$, $LH$, and split-off band) [Fig.~4(a)].  In wz-InN and wz-GaN, by contrast, the already existing splitting between the two top VBs---the $A$ and the $B$ band---does not significantly increase if similar strain is applied [Fig.~4(b,c)].  Note, that this is true only for \emph{compressive} biaxial strain in the basal plane.  As a result, the confined $A$- and $B$-type hole states in InGaN QDs are energetically close. We demonstrated recently \cite{winkelnkemper2006}, that the confined hole ground state ($h_A$) is predominantly of $A$-type, while the first excited hole state ($h_B$) is predominantly of $B$-type.  Both have an $s$-type envelope function and are separated by $\approx7$\,meV.  We will show in the following that the transitions between the electron ground state ($e_0$) and both of these hole states have substantial oscillator strengths and that even a small structural anisotropy of the QDs leads to pronounced polarization of these transitions in orthogonal directions.

For the electronic structure calculations, we use a 3-dimensional strain-dependent eight-band $\kpe$ model implemented on a finite differences grid.  The model includes piezoelectric and pyroelectric effects as well as crystal-field splitting and spin-orbit interaction.  Few-particle states are calculated using a self-consistent Hartree (mean field) approach.  The entire method is described in detail elsewhere \cite{winkelnkemper2006}.  The current parameter set has been adjusted \cite{footnote_winkseg} to include recently developed band-dispersion parameters \cite{rinke2006} derived from accurate quasiparticle energy calculations  \cite{rinke2005}. Size and composition of the QDs has been derived from HRTEM measurements which suggest a spinodal decomposition of the InGaN layer, thus forming fluctuation-induced QDs \cite{gerthsen2003,seguin2004}. In our model we assume an ellipsoid with graded In concentration $x$, with  $x_\tn{c}=50$\,\% at the QD's center and $x_\tn{e}=5$\,\% at its edges.  The QD has a height of $d_\tn{z}=2.0$\,nm (parallel to the $c$-axis), a lateral diameter of $d_\tn{x}=d_\tn{y}=5.2$\,nm (in the basal plane), and is embedded in an InGaN quantum well with a height of $2.0$\,nm and In concentration of $x_\tn{w}=x_\tn{e}=5$\,\%.

\begin{figure}[t] 
  \includegraphics[clip,width=0.9\columnwidth]{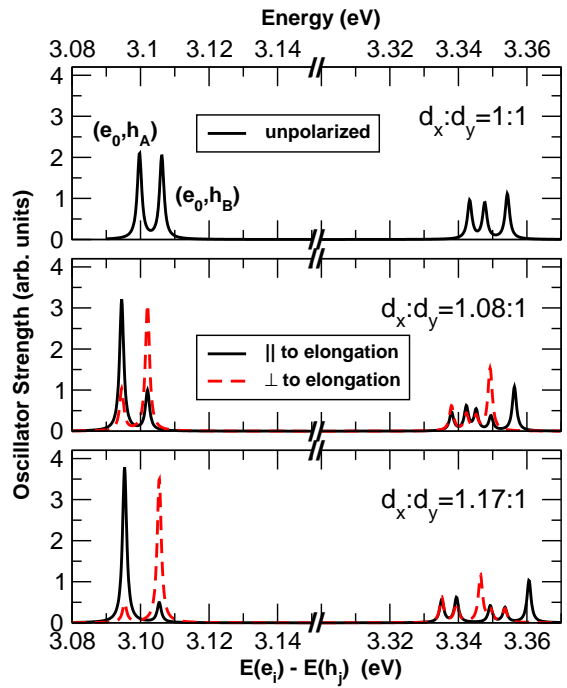}
  \caption{\label{fig:single_osc}Oscillator strengths between electron and hole states in QDs with different degrees of elongation. The solid (dashed) lines show the oscillator strength parallel (perpendicular) to the QD elongation. The $x$ axis shows the energy difference between the involved electron and hole states.}
\end{figure}

We calculated the oscillator strengths between the single-particle electron and hole states as a function of anisotropy in the basal plane (Fig.~5).  The in-plane aspect ratio of the QD, $d_\tn{x}$:$d_\tn{y}$, has been altered between $1$:$1$ and $1.25$:$1$, while the total amount of indium has been kept constant.  Three bound electron levels and a much larger number of bound hole levels ($>$$6$) are found.  The spectra in Fig.~5 include all single-particle (electron-hole-pair) transitions involving one of the three bound electron levels and one of the first six bound hole levels.  Non-vanishing oscillator strengths were found for the transition between the electron ground state and the hole ground state ($e_0$,$h_A$), the transition between the electron ground state and the first excited hole state ($e_0$,$h_B$), and transitions involving higher excited hole states and excited electron states. However, the latter have much higher transition energies and weaker oscillator strengths (Fig.~5).  The ($e_0$,$h_A$) and ($e_0$,$h_B$) transitions are energetically separated by only $7-10$\,meV.  Both transition lines are unpolarized for round QDs ($d_\tn{x}$:$d_\tn{y}=1$:$1$).  For asymmetric QDs the $A$-band transition is linearly polarized parallel to the QDs' long axis and the $B$-band transition perpendicular to it.  Even for a very slight asymmetry of $1.08$:$1$, the degree of polarization of both lines is already $\approx 3$:$1$. An elongation of $1.25$:$1$ results in an almost complete linear polarization ($\approx 20$:$1$).

These two electron-hole-pair configurations [($e_0$,$h_A$) and ($e_0$,$h_B$)] take part in many different confined few-particle complexes, such as excitons (X), biexcitons (XX), or charged excitonic complexes. Thus, the emission spectra of these complexes show the same polarization behavior---linear, in orthogonal directions---as found for the single particle transitions (Fig.~6). Considering the irregular shapes of the fluctuation-induced QDs, an elongation in the basal plane is very likely.  We therefore conclude, that the participation of $A$- and $B$-band transitions is the reason for the observed polarization of the emission lines. The cause for the regular orientation of the QDs along [$11\overline{2}0$] and/or [$\overline{1}100$] remains unclear.

\begin{figure}[t] 
  \includegraphics[clip,width=0.9\columnwidth]{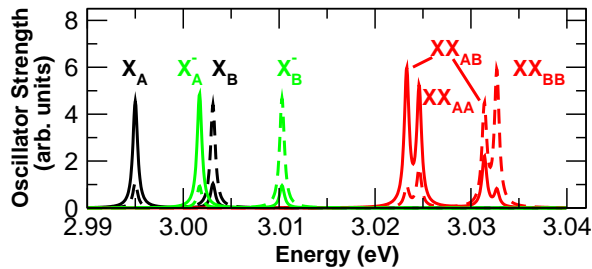}
  \caption{\label{fig:fewpart_osc}Calculated few-particle transitions for a QD with an in-plane elongation of $d_\tn{x}$:$d_\tn{y}=1.08$:$1$; $X$ (black lines), $XX$ (red lines), and $X^{\tn{-}}$ (green lines) lines. The solid (dashed) lines show the oscillator strength parallel (perpendicular) to the QD elongation.}
\label{figure6}\end{figure}

As an example of a possible few-particle spectrum, Fig.~6 shows the calculated oscillator strengths for the confined $X$, $XX$, and negative trion, $X^\tn{-}$, including both, the $A$- and $B$-like states.  All transition lines are linearly polarized.  The polarization direction of each transition depends on whether an $A$ or $B$ hole state is involved in the recombination process. The resulting spectrum is qualitatively similar to the measured spectra.  We would like to point out, that the exact transition energy of each line depends heavily on a number of parameters which are yet unknown with sufficient precision, most important the crystal-field and spin-orbit splitting energies of InGaN, the strength of the built-in electric fields, and the exact shape and size of the QDs. Also, the occupation probabilities of the different states are unknown.  Therefore, the emission lines in Fig.~2 can not be assigned to specific excitonic complexes yet.

In conclusion we have reported single-QD emission spectra from InGaN QDs with up to five lines per QD. The emission lines are linearly polarized in orthogonal directions.  The large number of emission lines has been explained with the existence of higher excitonic complexes; their polarization with  a slight anisotropy of the QDs in the basal plane and recombinations involving hole states which are either formed by the $A$ or $B$ valence band. 

This work was in part funded by Sfb 296 of DFG and SANDiE Network of Excellence (No.\ NMP4-CT-2004-500101) of the European Commission. Parts of the calculations were performed on the IBM pSeries 690 supercomputer at HLRN within Project No.\ bep00014.

\end{document}